\documentstyle[fleqn,amssymb,psfig]{tp}

\newcommand\ba{\begin{array}}
\newcommand\ea{\end{array}}
\newcommand\bc{\begin{center}}
\newcommand\ec{\end{center}}
\newcommand\be{\begin{enumerate}}  
\newcommand\ee{\end{enumerate}}  
\newcommand\bi{\begin{itemize}}  
\newcommand\ei{\end{itemize}}  
\newcommand\bd{\begin{description}}  
\newcommand\ed{\end{description}}  
\newcommand\beq{\begin{equation}}  
\newcommand\eeq{\end{equation}}  
\newcommand\beqa{\begin{eqnarray*}}  
\newcommand\eeqa{\end{eqnarray*}}

\newcommand{\etal}{{\em et al.\ }}

\newcommand\mathC{\mkern1mu\raise2.2pt\hbox{$\scriptscriptstyle|$}
                {\mkern-7mu\rm C}}

\def\exl{\raise1pt\hbox{$\scriptstyle<$}}
\def\exr{\raise1pt\hbox{$\,\scriptstyle>$}}

\begin{document}

\twocolumn[
\title{Analysis Issues for Large CMB Data Sets}
\author{Krzysztof M. G\'orski$^{(1,2)}$,\\
E. Hivon$^{(1)}$, and B.D. Wandelt$^{(1)}$\\
{\it $^{(1)}$TAC, Juliane Maries Vej 30, DK-2100 Copenhagen, Denmark}\\
on leave from \it{ $^{(2)}$Warsaw University Observatory, Warsaw, Poland}\\
}
\vspace*{16pt}   

ABSTRACT.\
Multi-frequency, high resolution, full sky measurements of the anisotropy in
both temperature and polarisation of the cosmic microwave
background radiation  
are the goals of the satellite missions MAP (NASA) and Planck (ESA).
The ultimate data products of these missions ---
multiple microwave sky maps, each of which will have to comprise 
more than $\sim 10^6$ pixels in order to render the angular 
resolution of the instruments ---
will present serious challenges to those involved in the
analysis and scientific exploitation of the results of both surveys.
Some considerations of the relevant aspects of the mathematical structure
of future CMB data sets are presented in this contribution.

\endabstract]

\markboth{Krzysztof M. G\'orski \etal}{Analysis Issues for Large CMB Data Sets}

\small


\section{Introduction}

The extraordinary success of NASA's  {\it COBE} satellite mission 
(see www.gsfc.nasa.gov/astro/cobe),
and in particular the {\it COBE}-DMR discovery of anisotropy in the temperature 
of the cosmic microwave background (CMB) radiation (Smoot et al. 1992), has transformed 
the field of experimental and theoretical studies of the CMB as an astronomical
window on the early universe
into one of the most vibrant areas of modern cosmology.
The explosive growth of activity in the CMB community revealed the promise of
dramatic improvement in our knowledge of the early universe 
that should materialize given the continuation of  efforts to study
the CMB from space.
This, and the realization of a sufficient improvement in microwave detector technology
suitable for space experiments, 
resulted in both NASA and the European Space Agency approving proposals 
to conduct the next generation satellite missions to study the CMB:
MAP (see  http://map.gsfc.nasa.gov) to be launched in 2000, and Planck
(see http://astro.estec.esa.nl/SA-general/Projects/Planck) to be launched in 2007.

Both MAP and Planck were designed to surpass the {\it COBE}-DMR capabilities to measure
anisotropy in the temperature and polarization of the CMB by orders of magnitude.
Both missions will observe the CMB from the Earth-Sun L-2 
point using $\sim 1.5$m telescopes.
MAP will use passively cooled HEMT detectors with five frequency bands between
22 and 90 GHz, and will reach an angular resolution between 0.93 and 0.21 deg,
respectively, and is expected to render a noise sensitivity 
of $\sim 20 \mu$K per 0.3 x 0.3 degree pixel
achieved by combining the three highest frequency channels.
Planck will comprise two instruments involving complementary detector technologies
to yield unprecedented, very wide frequency coverage.
The Low Frequency Instrument will use an actively cooled, 
tuned radio receiver array operated at 20 K with four frequency bands between 30 and 100
GHz, should achieve an angular resolution between 33 and 10 arcmin, respectively,
and should render a noise sensitivity per resolution element between 4 and 12 $\mu$K, 
respectively.
The High Frequency Instrument will use bolometer arrays operated at 0.1 K  with 
six frequency bands between 100 and 857 GHz, should achieve an angular resolution
between 10  
and 5 arcmin, respectively, and should render a noise sensitivity per resolution element
between 4.6 and 12 $\mu$K in the  CMB dominated channels (100-217 GHz).

It is widely recognized that if both missions succeed in achieving these spectacular 
specifications during observation of the CMB sky our knowledge of the universe
will be furthered dramatically (see the contributions by C. Lawrence and 
N. Sugiyama, and references therein, in these proceedings). 
Indeed, it is expected that the results of MAP and Planck
should allow us to answer decisively both 
fundamental questions about the global
properties of the universe --- 
the average density of its matter content, the expansion rate,
global curvature, the existence of a cosmological constant and/or 
a cosmological background of gravity waves, to name a few ---
and to unravel the properties  of a number of astronomical
objects --- for example the detailed picture of emission from our Galaxy, and 
infrared emission from distant galaxies and galaxy clusters.
 
At the same time it should be realized that the data products of the future CMB missions
will be by no means trivial to work with. The enormity of the expected
scientific return will 
come at the price of a concerted effort to improve all currently available methods 
of analysis of very large CMB data sets, which, in order to answer properly 
some of the grand questions, will need to be studied globally.
 
As we have learned working with the {\it COBE} mission products, the digitized
sky map is an essential intermediate 
stage in information processing between the entry point of data acquisition by the 
instruments --- very large time ordered data streams,
and the final stage of astrophysical analysis --- 
typically producing a `few' numerical values
of physical parameters of interest. 
{\it COBE}-DMR sky maps (three frequency bands, two channels each, 6144 pixels per map)
were considered large at the time of their release.
So, looking forward to the giant leap to be taken by MAP and Planck in
mapping the entire CMB sky,
what do we mean when saying `very large future CMB data sets'?
Very simply, if we focus on the case of whole sky CMB measurements 
at the angular resolution
of $\sim 10$ arcmin (FWHM), and presume that a few pixels per resolution element 
should be used to discretise the signal in a 
non-damaging way (i.e. in such a way that discretisation effects are sufficiently 
sub-dominant with respect to the effects of instrument response),
we require that the entire map should comprise at least 
$N_{pix}\sim $ a few $\times 1.5\, 10^6$ 
pixels.
More pixels than that will be needed to represent the Planck-HFI higher resolution channels.
This estimate, $N_{pix}$, should then be multiplied by  the number of frequency bands 
(or, indeed, by the number of individual
observing channels --- 74 in the case of Planck --- for the analysis work to be done
before the
final coadded maps are made for each frequency band) to render an approximate expected
size of the already compressed form of data, which would 
be the input to the astrophysical analysis pipeline. 
Clearly, it is easy to end up with an estimate of many GBy. Hence it appears that
some attention should be given to devising such high resolution CMB map
structures which would maximally facilitate future analyses of large size data sets.

The main part of this contribution is devoted to a presentation of the
properties of one such proposed approach to high resolution full sky 
map making --- the Hierarchical
Equal Area and isoLatitutde Pixelization of the sphere (HEALPIX, see 
http://www.tac.dk/$\sim$healpix). The remaining part of this paper is a digression on the
issue of the (non-)gaussianity of the CMB sky.

\section{Discretisation of Functions on a Sphere with HEALPIX for 
High Resolution Applications}

Operations which are, or will be, routinely executed in the analysis of CMB maps include
convolutions with local and global kernels, Fourier analysis with spherical harmonics
and power spectrum estimation,
wavelet decomposition, nearest-neighbour searches, topological analysis 
(minimum/maximum search, Minkowski functionals,
etc.), and more. Some of these operations become prohibitively slow  
if the sampling of functions on a sphere, and the related structure of 
the discrete data set, is not designed carefully. 

Typically, a whole sky map rendered by a CMB experiment contains the sky signals,
which should be (by design) strongly band-width limited (in the sense of spatial Fourier
decompostion) by the instrument's response 
function, and projection of the instrumental noise, which, at 
\begin{figure*}[t]
\centering\mbox{\psfig{figure=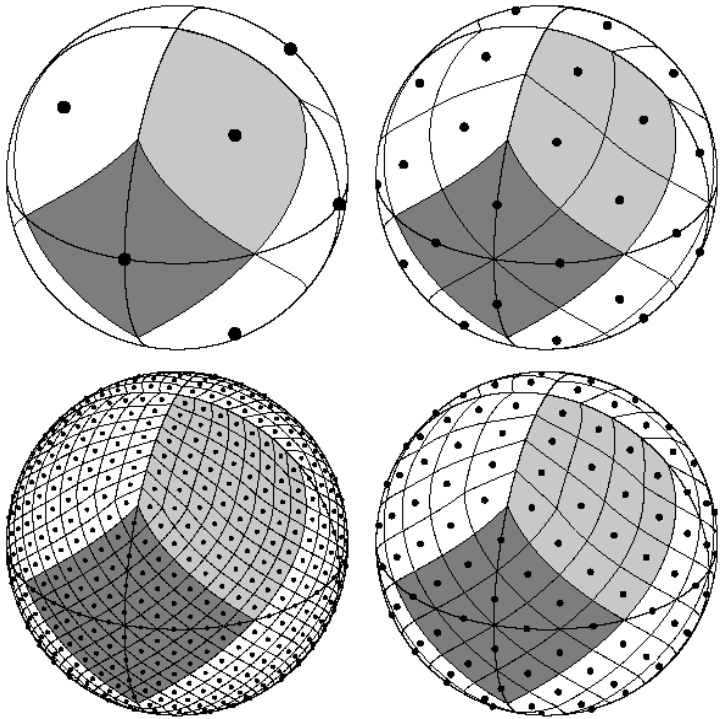,height=9.5cm}}
\caption[]{Orthographic view of HEALPIX division of a sphere. 
Overplot of equator and  meridians illustrates  octahedral symmetry of the 
HEALPIX construction. 
Light-gray shading shows one of eight (four north, and four south) identical polar 
base-resolution pixels. 
Dark-gray shading shows one of four identical equatorial base-resolution pixels. 
Moving clockwise from the upper left 
panel the grid is hierarchically subdivided with 
the grid resolution parameter equal to $N_{side} = \,1,\,2,\,4,\,8$, 
and the total number of pixels  equal to 
$N_{pix} = 12 \times N_{side}^2 = \,12,\,48,\,192,\,768$. 
All pixel centers are located on $N_{ring} = 4 \times N_{side} - 1$ rings of 
constant latitude.
Within each panel the areas of all pixels are identical.}
\label{HEALPIX}
\end{figure*}
least near the discretisation 
scale,  should be random, with a band-width significantly exceeding that of all the 
signals. 

With all these considerations in mind one may propose the following list of desiderata 
for the mathematical structure of discrete whole sky maps:

1. {\bf Hierarchical structure of the data base}. This is recognized as 
essential for very large data bases, and was indeed postulated already in construction
of the Quadrilateralized Spherical Cube 
(see http://www.gsfc.nasa.gov/ast\-ro/co\-be/skymap\_info.html), 
which was used for all the
{\it COBE} data. A simple argument in favour of this states that the data elements  
which are nearby in a multi-dimensional configuration space (here, on the surface of 
a sphere), are also nearby in the tree structure of the data base. This property 
facilitates various topological methods of analysis, and allows for easy construction
of wavelet transforms on triangular and quadrilateral grids.

2. {\bf Equal areas of discrete elements of partition}. This is advantageous because
white noise at the sampling frequency of the instrument gets integrated exactly into
white noise in the pixel space, and sky signals are sampled without regional dependence
(although still care must be taken to choose a pixel size
sufficiently small compared to the 
instrumental resolution to avoid excessive, and pixel shape dependent, signal
smoothing).

3. {\bf Iso-Latitude distribution of discrete area elements on a sphere}.  This property
is essential for computational speed in all operations involving evaluations of spherical
harmonics. Since the associated Legendre polynomials are evaluated via
slow recursions, any sampling grid deviations from an iso-latitude
distribution result in a prohibitive loss of computational performance
with the growing number of sampling points.

Various known sampling distributions on a sphere fail to meet simultaneously 
all of these requirements:

i) Quadrilateralized Spherical Cube obeys points 1 and
(appro\-xi\-ma\-te\-ly) 2, but fails
on point 3, and cannot be used for efficient Fourier analysis at high resolution.

ii) Equidistant Cylindrical Projection, a very common computational tool in geophysics,
and climate modeling, and recently implemented for CMB work 
(Muciaccia, Natoli, Vittorio, 1998), satisfies points 1 and 3, but by construction
fails with point 2. This is a nuisance from the point of view of application 
to full sky survey data due to wasteful oversampling near the poles of the map.

iii) Hexagonal sampling grids with icosahedral symmetry 
perform  superbly in those applications where near uniformity
of sampling on a sphere is essential (Saff, Kuijlaars, 1997),
and can be devised to meet requirement 2 (see e.g. Tegmark 1996). However, by
construction they fail {\it both} points 1 and 3.

iv) Igloo-type constructions are devised to satisfy point 3
(E. Wright, 1997, private communication; Crittenden \& Turok, 1998). 
Point 2 can be satisfied to reasonable accuracy if quite a large
number of base-resolution pixels is used, which precludes
the efficient construction of simple wavelet transforms. 
Conversely, a tree-structure seeded with a small number of 
base-resolution pixels forces significant variations in both 
area and shape of the pixels. 

All three requirements formulated above are satisfied by construction with the
Hierarchical Equal Area, iso-Latitude Pixelisation (HEALPIX) 
of the sphere (G\'orski 1998, Hivon, G\'orski, 1998), which is shown in Figure 1. 

HEALPIX is a genuinely curvilinear partition of the sphere into exactly equal area
quadrilaterals of varying shape. The base-resolution comprises twelve pixels in three
rings around the poles and equator. 

The resolution of the grid is expressed by parameter $N_{side}$ which defines the number
of divisions along the side of a base-resolution pixel that is needed to reach a desired
high-resolution partition.

All pixel centers are placed on rings of constant latitude, 
and are equidistant in azimuth
(on each ring). All iso-latitude rings located between the upper and lower corners of
the equatorial base-resolution pixels, or in the equatorial zone, 
are divided into the same number of pixels: 
$N_{eq}= 4\times N_{side}$. The remaining rings are located within the
polar cap regions and contain a varying number of pixels, increasing 
from ring to ring, with increasing distance
from the poles, by one pixel within each quadrant. 

Pixel boundaries are non-geodesic and take the very simple 
form: $\cos \theta = a + b \times \phi$ in the equatorial zone, 
and $\cos \theta = a + b / \phi^2$ in the polar caps. 
This allows one to explicitly check by simple analytical integration the 
exact area equality among pixels,
and to compute efficiently more complex objects, 
e.g. the Fourier transforms of individual pixels.

\begin{figure*} [t]
\centering\mbox{\psfig{figure=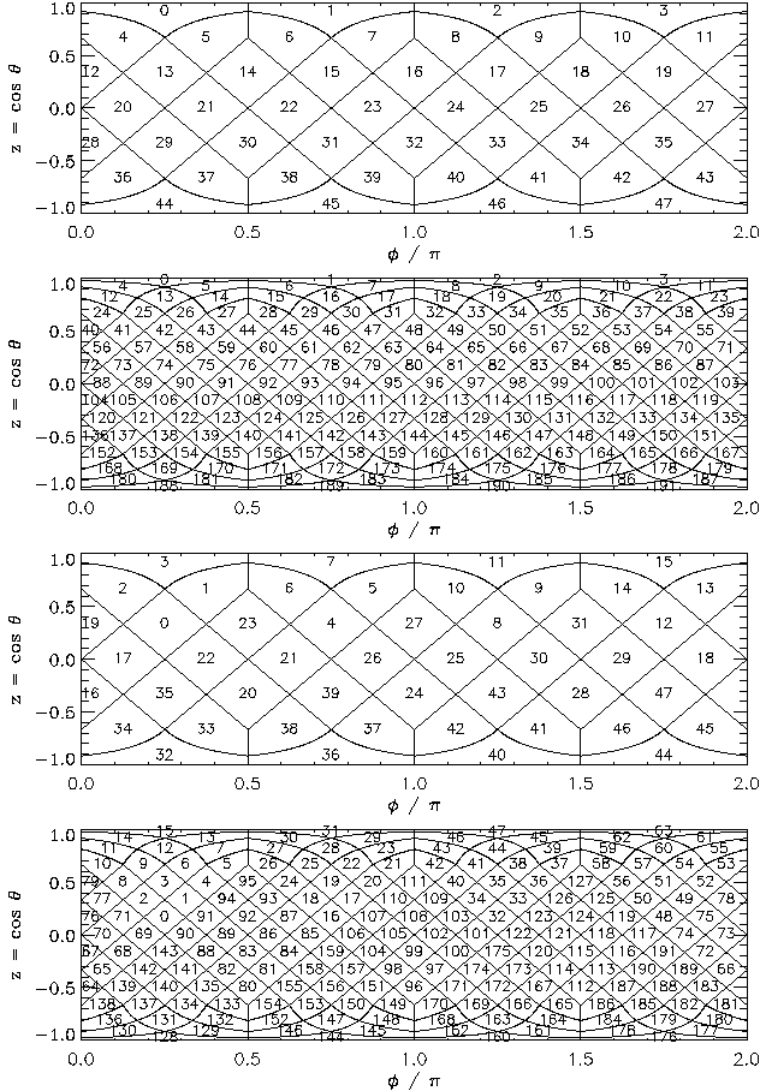,height=14.5cm}}
\caption[]{Cylindrical projection of the HEALPIX  division of a
sphere and two natural pixel numbering schemes (ring and nested) 
allowed by HEALPIX. Both numbering schemes map the two dimensional distribution
of discrete area elements on a sphere into the one dimensional, integer pixel number array,
which is essential for computations involving data sets with very large total pixel numbers.
>From top to bottom:
Panel one (resolution parameter $N_{side} = 2$) and panel two ($N_{side} = 4$)
show the ring scheme for pixel numbering, with the pixel number winding 
down from north to south pole through the consecutive isolatitude rings.
Panel three (resolution parameter $N_{side} = 2$) and panel four ($N_{side} = 4$)
show the nested scheme for pixel numbering within which the pixel number grows
with consecutive hierarchical subdivisions on a tree structure seeded by the twelve 
base-resolution pixels. 
}
\label{Numbering}
\end{figure*}

Specific geometrical properties allow HEALPIX to support two different
numbering schemes for the pixels, as illustrated in the Figure 2.
First, one can simply count the pixels moving down from the north 
to the south pole along each
iso-latitude ring. It is in this scheme that Fourier transforms with spherical harmonics
are easy to implement.
Second, one can replicate the tree structure of pixel numbering used e.g. with
the Quadrilateralized Spherical Cube. This can easily be implemented
since, due to the simple 
description of pixel boundaries, the analytical mapping of the HEALPIX
base-resolution elements (curvilinear
quadrilaterals) into a [0,1]$\times$[0,1] square exists.
This tree structure, a.k.a. nested scheme, allows one to implement efficiently all
applications involving  nearest-neighbour searches
(see Wandelt, Hivon, and G\'orski 1998),
and also allows for an immediate
construction of the fast Haar wavelet transform on HEALPIX. 

We have developed a HEALPIX software package which contains a suite of programs to
simulate and analyse full sky CMB maps, display the results, manipulate the
transformations from physical space to pixel number (and the reverse),
in both numbering schemes,
and conduct nearest-neighbour and maxima/minima searches on the maps.
The package is available to the public  at  http://www.tac.dk/$\sim$healpix. 
HEALPIX is presently the format chosen by the MAP 
collaboration to be used for the production
of sky maps from the mission data 
(see http://map.gsfc.nasa.gov/html/tech\-nical\_info.html). 
HEALPIX software is widely used 
for simulation work within both LFI and HFI consortia of Planck collaboration. 

\section{Digression: Non-Gaussianity in the DMR data?}

The remainder of this contribution is a digression, which was stimulated by 
questions during the presentation. 

In a recent paper (Ferreira, Magueijo, and G\'orski, 1998) the authors 
argued that an estimate of the reduced bi-spectrum of the CMB
anisotropy in the {\it COBE}-DMR 53 and 90 GHz 4yr data is significantly
deviant from the distribution simulated from a Gaussian ensemble.
Their work addressed the possibility of the effect being due to foreground 
emission from our Galaxy, but failed to explain it as originating from
the known distributions of galactic emission at low and high
frequencies (as traced by radio synchrotron and dust maps respectively).

An update on the status of our (FMG) ongoing tests and searches for alternatives 
tends to the conclusion
that the CMB itself may have revealed its non-Gaussian
nature. Evidence in support of this is as follows:
1) the reduced bi-spectrum estimated on the DMR difference maps (noise-only combinations
of the DMR sky maps) shows {\it no} excess at the conspicuous 
multipole index of $\ell =16$
where the dominant contribution to the original effect was found;
2) likewise, the reduced bi-spectrum estimated on a suite of 
systematic effect model templates
that were derived during  the {\it COBE}-DMR mission, turned out to be insignificant in 
amplitude --- itself a testimony to the superb systematic 
effect control in the DMR data analysis ---
and again, quite unlike the original effect in the multipole index dependence.

Hence, our (FMG) original conclusions remain
unaffected: unless an as yet completely
unrecognized foreground emission results in the detected non-Gaussianity, the remaining
candidate for the source of this effect is the CMB anisotropy itself.

\section*{Acknowledgments}

The work of KMG, EH, and BDW was funded  by the Dansk Grundforskningsfond through its
funding for TAC.



\end{document}